\documentclass[runningheads]{llncs}
\usepackage{graphicx}
\usepackage{wrapfig}
\usepackage{float}
\usepackage{amsmath}
\usepackage{amssymb}
\usepackage{mathtools}
\usepackage[english]{babel}
\usepackage{cite}
\usepackage{textcmds}
\usepackage[utf8]{inputenc}
\usepackage{lineno}

\usepackage{xspace}
\usepackage{dirtytalk}

\newcommand{\Heraklit}{\textsc{Heraklit}\xspace}

\DeclareMathOperator{\compose}{{{\normalfont\text{\textbullet}}}}

\DeclareMathOperator{\elm}{\textit{elm}}

\DeclareMathOperator{\sthinkingphilosophers}{\emph{thinking philosophers}}
\DeclareMathOperator{\savailableforks}{\emph{available forks}}

\begin{document}
\title{Essentials of Petri nets}
%
%

\author{Wolfgang Reisig\inst{3}\orcidID{0000-0002-7026-2810} \and Peter Fettke\inst{1,2}\orcidID{0000-0002-0624-4431}
}
\authorrunning{W. Reisig, P. Fettke}
%

\institute{German Research Center for Artificial Intelligence (DFKI), Saarbr\"ucken, Germany \\
\email{peter.fettke@dfki.de}\\ \and
Saarland University, Saarbr\"ucken, Germany \\ \and
Humboldt-Universität zu Berlin, Berlin, Germany \\ 
\email{reisig@informatik.hu-berlin.de}}

\maketitle 
\begin{abstract}
This contribution highlights some concepts and aspects of Petri nets that are frequently neglected, but that the authors consider important or interesting, or that Carl Adam Petri emphasized.

\end{abstract}

\section*{Introduction}\label{sec:intro}

This contribution presents a collection of concepts and aspects of Petri nets, that fundamentally differ from other models of computation, and in particular from other models of concurrency. As a previous staff member of Carl Adam Petri, the first author of this contribution, Wolfgang, tried to select topics that he assumes Petri would have liked to see in such a collection (among many others, not addressed here). Nevertheless, some views and aspects of this contribution are personal perception of today, looking back 40 years. In the early 1980s, other themes would certainly have been chosen.

We start trying to give an account of the technical circumstances of the emerging computer science in the late 1950s and early 1960s, when Petri submitted his seminal PhD thesis. In seven temporal steps we present some highlights of concepts, as they developed until today. 

The reader is assumed to be familiar with the standard notions and graphical conventions of Petri nets \cite{Reisig_13}. However, he or she must be prepared to accept slightly different views.

\section{The late 1950s}\label{sec:1}
In the 1950s, computing started to become scientifically and commercially interesting. In those times, computers weighted tons and costed millions of dollars. Theoretical Computer Science was emerging. Undisputedly, Turing machines were conceived as adequate abstract models for computers. First papers on automata theory and formal languages emerged.

\section{The early 1960s}\label{sec:2}
In this atmosphere, in 1962, Carl Adam Petri submitted a text, entitled “Communication with Automata”, as his PhD thesis \cite{Petri_62}. This title deliberately has two readings: first, a person communicates with an automaton, or second, a group of persons communicates with the help of automata. With this thesis, Petri intended to strive towards a comprehensive theory for discrete dynamic systems, not just another theory for symbol processing automata. The thesis itself is a collection of ideas, a research agenda, towards this goal.

Particularly striking is a nice thought experiment in the thesis: How to implement a non-primitive recursive function $f$? One may write a corresponding program $p$, and for a given argument $a$, one may run the program $p$ on a given computer $C$ and hope that the computation will stop, returning $f(a)$. But one can not be sure that the computation will eventually stop. And even if it will eventually stop, beforehand one cannot estimate the amount of resources needed. So, the resources of $C$, in particular storage for intermediate results, may not suffice, and the computation may eventually fail due to lack of resources of $C$. In this case, one may engage a bigger computer $C'$ with more resources, and start computation from scratch. Of course, also $C'$ may fail. One may iterate this process until either the resources suffice, or one gives up.

\begin{figure}[t]
\centering
\includegraphics[scale=.4]{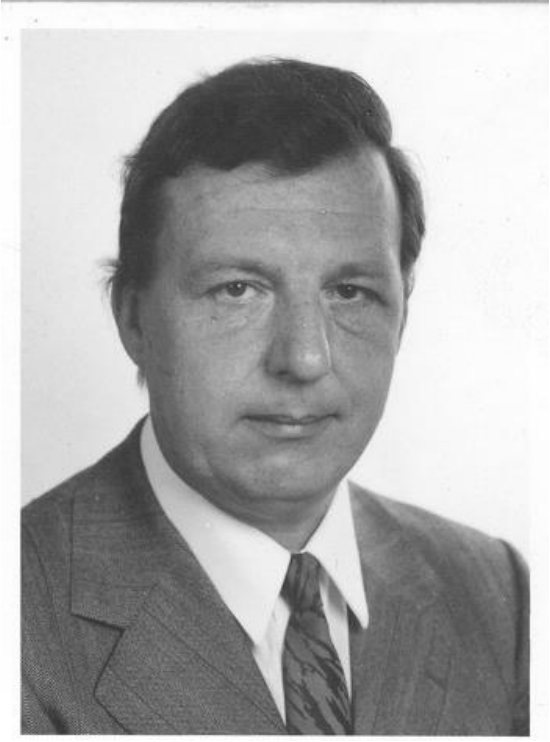}
\caption{Carl Adam Petri in the late 1950s}
\label{fig:01}
\end{figure}

Petri suggested to improve this procedure: Instead of replacing the computer $C$ by $C'$, just extend $C$ by more storage, and continue computing! Iterate this procedure until either the resources suffice, or give up. This, however, requires an architecture that can be extended unboundedly often. Can such an architecture physically be achieved? Petri argues that this requires the absence of any global feature, in particular of a global clock. Generally, any construction element must do with a bounded fan out of wires, and with a maximal length of all wires, to guarantee a stable clock pulse. This in turn prevents unbounded many extensions to be connected to one device. So, no device can directly be connected to all extensions. Summing up, this kind of architecture is feasible at the price of asynchronous components, and locally bounded causes and effects of events. 

This thought experiment can be interpreted as the observation that asynchronous systems can do more or be more efficient than synchronous systems. A corresponding theory of asynchronous system should reflect and quantify this observation. But such a theory is still missing. It would contradict, to some extent, the popular version of the Church thesis that everything that can be done with computers, can be mimicked by Turing machines. Petri's arguments did not meet the spirit of the time, and Petri's PhD thesis did not find much attention.

\section{The late 1960s}\label{sec:3}

As mentioned above, Petri intended to design a comprehensive theory for discrete dynamic systems in the real world, not just another theory for symbol processing automata. Therefore, Petri intended to base his theory on concepts of logic. 

The most fundamental notion of logic is the notion of a proposition or an assertion. Already Aristotle suggested this concept. A proposition is either always true (“5 is a prime number”) or always false (“The earth is flat”). Aristotle and his followers investigate special structures and combinations of propositions and their relation to aspects of the real world, until today (e.g. \cite{Suppes_57})

Petri suggested to parallel the “always true or always false” principle by “sometimes true and sometimes false”. As an example, “the bakery offers freshly baked bread” may now be true; it turns false by the bakery selling the last bread. It becomes true again, by baking and offering a new load of freshly baked bread. Hence, the truth value of such a proposition can occasionally be updated. Such a proposition is denoted as a (local) state; the proposition is true in case the state has been reached, and the proposition is false in case the state has been abandoned. 

\begin{figure}[H]
\centering
\includegraphics[width=\textwidth]{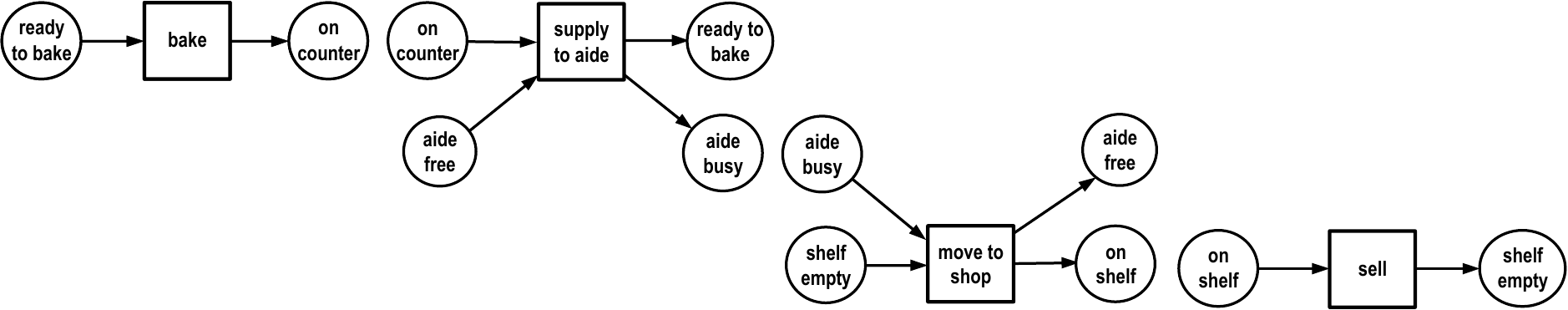}
\caption{Four steps of a bakery}
\label{fig:02}
\end{figure}

The question rises as how to organize updates of propositions. Petri suggested steps and their composition, resulting in runs. A step may occur; in this case, some previously reached states are abandoned, and some previously abandoned states are reached. Figure~\ref{fig:02} shows four steps: The leftmost, bake, can occur in case the local state ready to bake has been reached, and the local state on counter has been abandoned. Occurrence of bake then swaps the truth values of both local states. The step supply to aide apparently updates four states, in an obvious manner.

\begin{figure}[H]
\centering
\includegraphics[scale=.4]{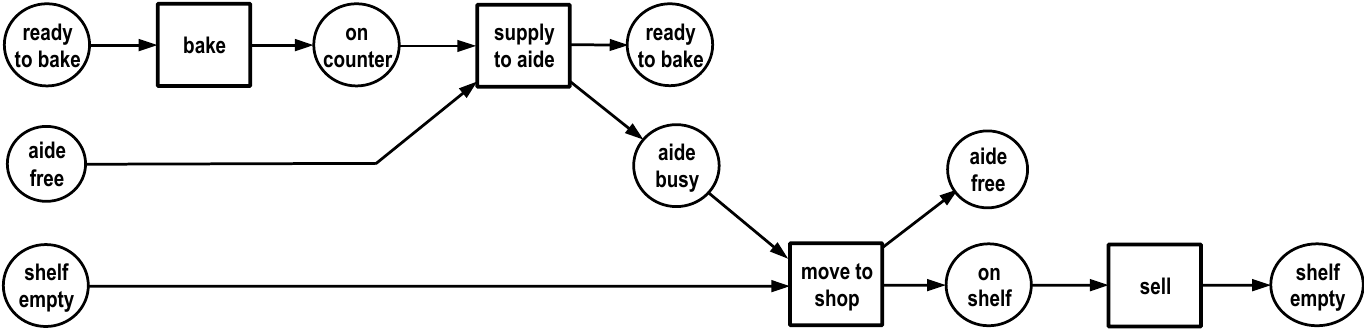}
\caption{A distributed run composed of four steps}
\label{fig:03}
\end{figure}

\begin{figure}[H]
\centering
\includegraphics[scale=.5]{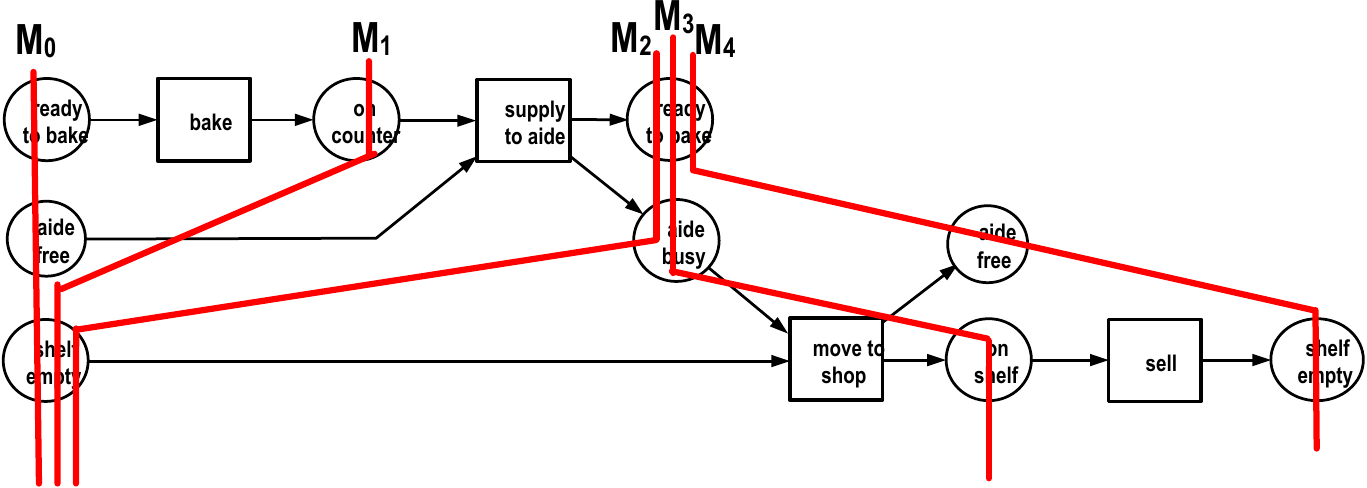}
\caption{Markings: global steps $M_0\xrightarrow[]{\text{bake}}M_1\xrightarrow[]{\text{supply to aide}}M_2\xrightarrow[]{\text{move to shop}}M_3\xrightarrow[]{\text{sell}}M_4$}
\label{fig:04}
\end{figure}

Figure~\ref{fig:02} outlines the four steps in such a way that identical states are drawn side by side. It is intuitively obvious to merge them, as in Figure~\ref{fig:03}, resulting in a distributed run. In such a run, an arc from an element $a$ to an element $b$ represents the causal dependency of $b$ from $a$.  So, this kind of runs present fine grained relationships between the elements of a run. However, it did not receive much attention at the time. It was quickly replaced by the global notion of markings, and global steps replaced local steps. Figure~\ref{fig:04} gives examples. We return to this issue later.

\section{The early 1970s}\label{sec:4}

In the early 1970s, two actual trends in informatics were particularly unfavorable for Petri Nets: Firstly, with the emerging notion of software engineering, it became common to construct software directly from intuitive imaginations, without too much of intermediate modeling. Secondly, automata theory became a success story, with deep rooted, interesting mathematical problems such as $P = {NP}$, complexity theory, and its relation to logic. To some extent, Petri nets adjusted to this trend: Each Petri net was assigned an automaton, its marking graph; classes of Petri net languages were identified; the reachability problem gained much attendance; etc. Much of this turned outside the area of notions and concepts that Petri envisaged as central for his theory.
Nevertheless, in the early 1970s, specific features for Petri nets were identified. This includes net classes such as free choice nets \cite{Hack_72}, and invariants \cite{Lautenbach_73}. In particular, place invariants exploit the observation that Petri net steps, in contrast to assignment statements of programming languages, are reversible, i.e., in the notation framework of vectors and matrices, with markings $M$ and $M'$ and a transition $t$, a step from $M$ to $M'$ can be written as: 

\begin{equation}\label{eq:01}
M' = M + \underline{t}.
\end{equation}

\noindent Equation \ref{eq:01} implies:
\begin{equation}
M = M' - \underline{t}.
\end{equation}

\begin{figure}[H]
\centering
\includegraphics[scale=.4]{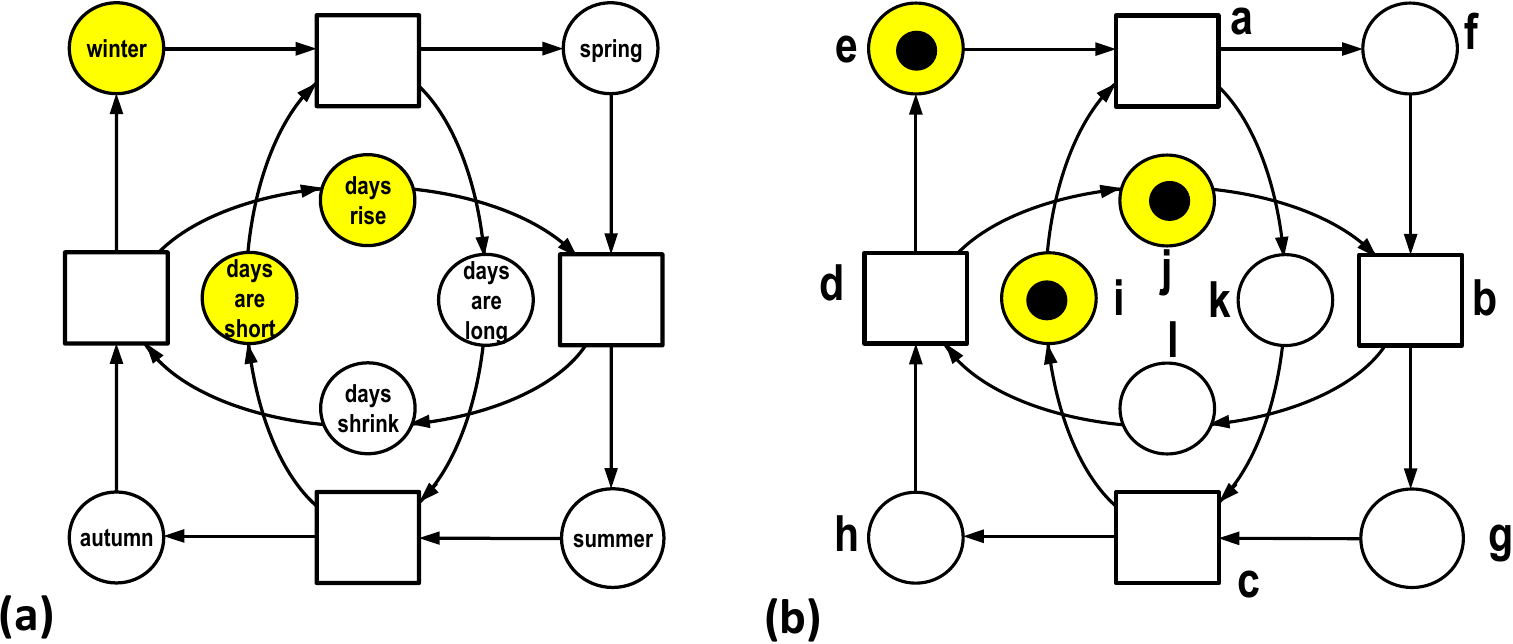}
\caption{Four seasons, intuitive model (a) and technical model (b)}
\label{fig:05}
\end{figure}

In this era, Petri suggested the four seasons model as particularly interesting \cite{Petri_79b}. Figure~\ref{fig:05}(a) shows this net. One may conceive it as intuitively impressive. But technically it is quite simple: it is just a sequence of four transitions, occurring repeatedly. It has no alternative and no concurrent transitions; so, central expressive means of Petri nets are missing. Furthermore, the four “inner” propositions, “days are short”, ”days rise”, ”days are long” and ”days shrink” are redundant: without them, the behavior remains. In the sequel, we motivate why Petri found this net particularly interesting.

Petri intended to define a notion of concurrency. Concurrency is intended to be a relation between system elements. The four seasons system is supposed to be the smallest system where each two elements are connected by a sequence of concurrent elements. Here we define concurrency as follows:

\begin{enumerate}

\item Two propositions $p$ and $q$ are concurrent, if and only if 
whenever $p$ is reached, before $p$ is abandoned, also $q$ is reached, and 
whenever $q$ is reached, before $q$ is abandoned, also $p$ is reached.

\item A proposition $p$ and a transition $t$ are concurrent, if and only if each time when $t$ occurs, $p$ is reached.
\end{enumerate}

\begin{figure}[H]
\centering
\includegraphics[scale=.4]{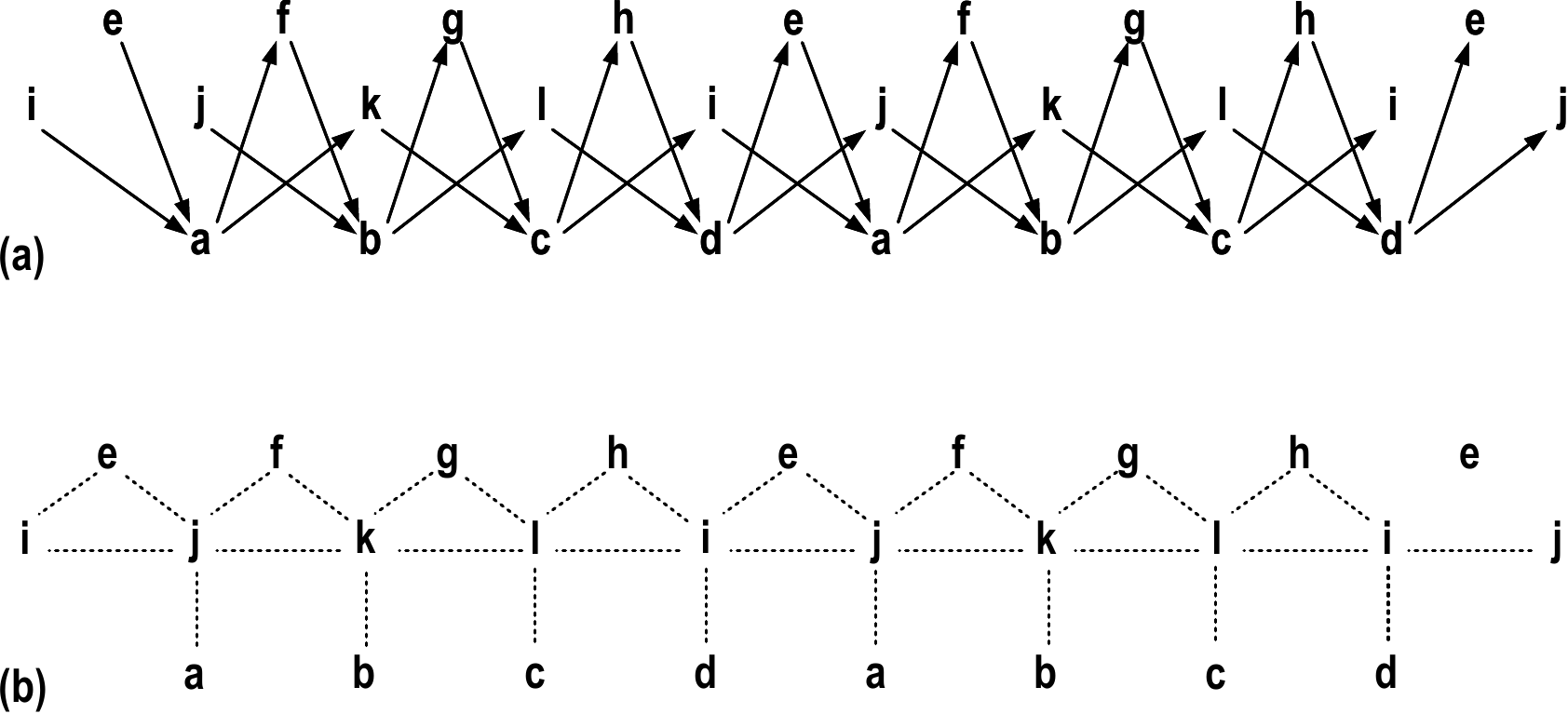}
\caption{Distributed run of the four seasons (a) and concurrency structure of the four season}
\label{fig:06}
\end{figure}

To exemplify this notion, as a technicality, Figure~\ref{fig:05}(b) renames the four season’s elements. Figure~\ref{fig:06}(a) shows a part of the distributed run of the four seasons system (as discussed in Section~3), albeit in confusing layout: the upper row includes the occurrences of the propositions of the “outer” circle, the middle row includes the occurrences of the propositions of the “inner” circle, and the lower row includes the occurrences of the transitions. This layout allows a quite symmetric representation of the concurrency structure of the distributed run, as in Figure~\ref{fig:06}(b): Concurrent pairs of elements are linked by a dotted line.

\begin{figure}[H]
\centering
\includegraphics[scale=.4]{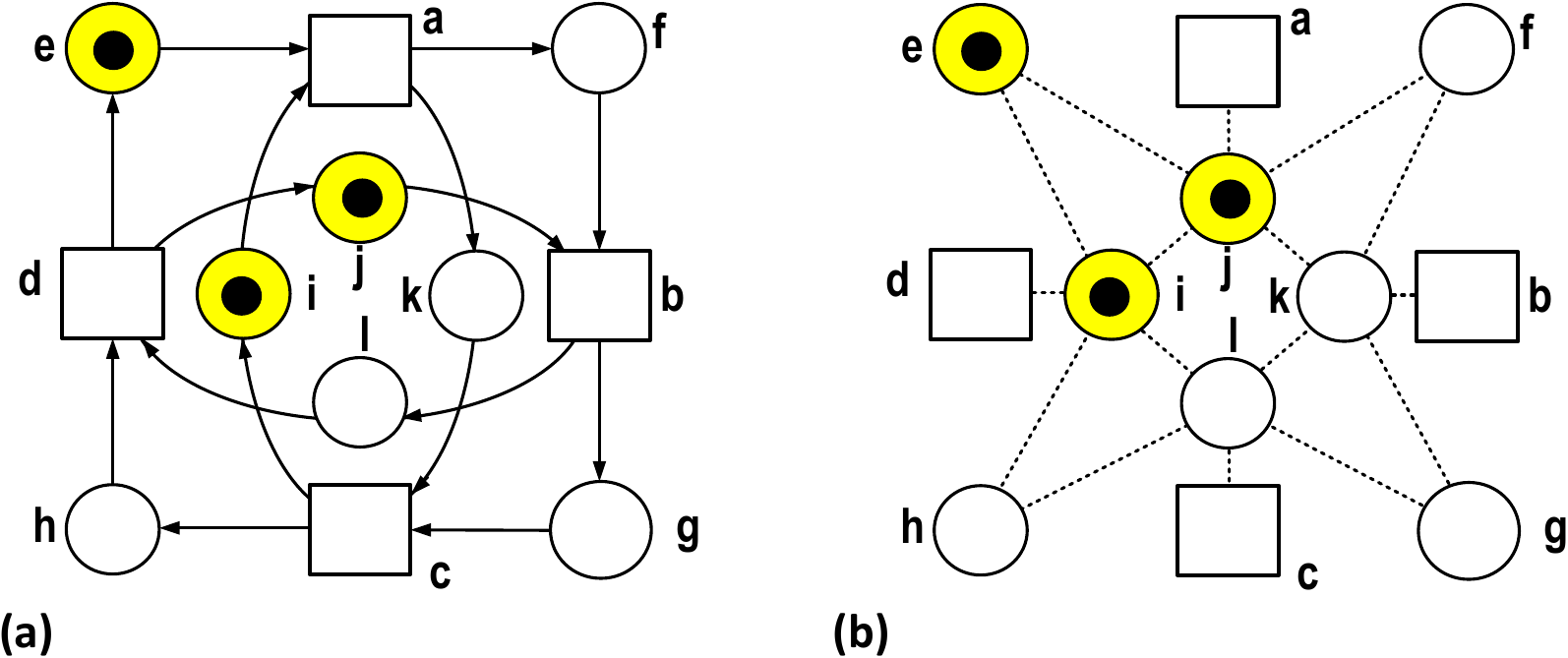}
\caption{Concurrency: (a) technical model, (b) concurrent model}
\label{fig:07}
\end{figure}

Figure~\ref{fig:07}(a) repeats the technical model of Figure~\ref{fig:06}, and Figure~\ref{fig:07}(b) shows the concurrency structure of the four seasons model. It is easy to see that from each place and each transition, each other net element is reachable by sequence of dotted lines.  Furthermore, skipping one of the propositions, would either ruin the behavior, or leave one of the transitions isolated.

This concurrency relation is intuitively appealing. But so far, this notion has not much impact in the world of concurrent systems.

\section{The late 1970s}\label{sec:5}

\begin{figure}[H]
\centering
\includegraphics[scale=.4]{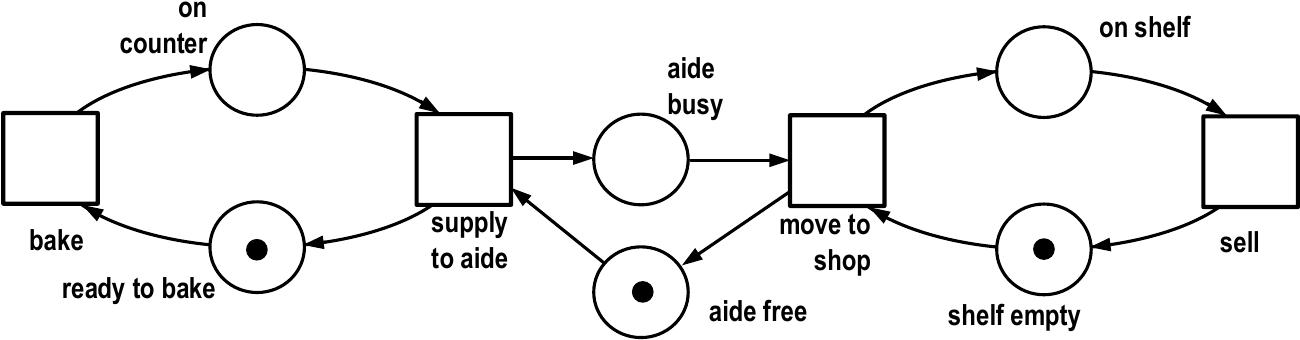}
\caption{Bakery}
\label{fig:08}
\end{figure}

\begin{figure}[H]
\centering
\includegraphics[scale=.4]{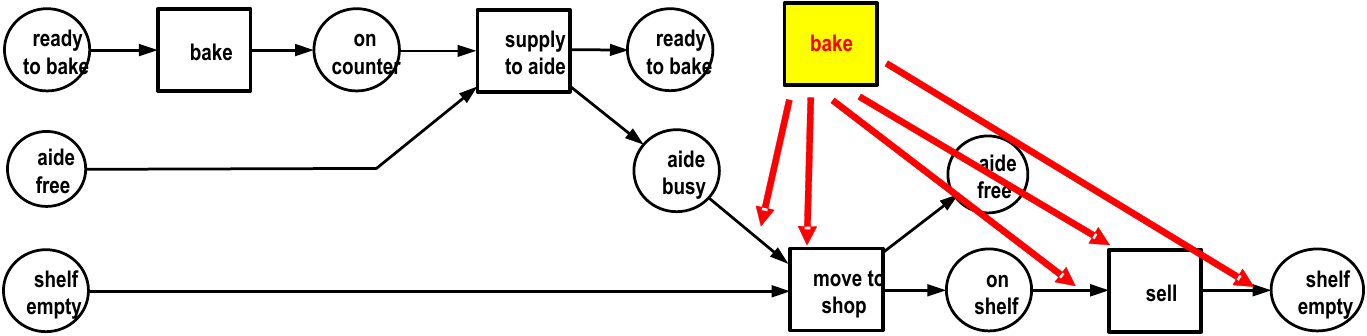}
\caption{Second bake}
\label{fig:09}
\end{figure}

\begin{figure}[H]
\centering
\includegraphics[scale=.4]{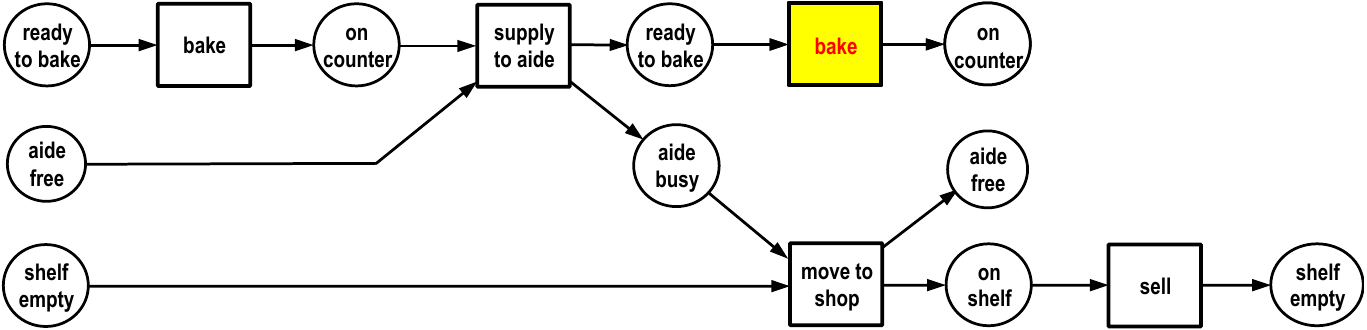}
\caption{Partial order}
\label{fig:10}
\end{figure}

\begin{figure}[H]
\centering
\includegraphics[width=\textwidth]{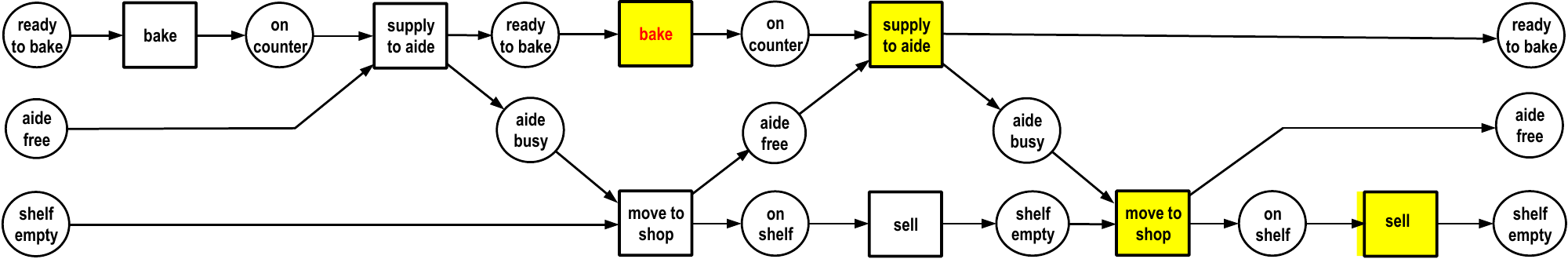}
\caption{2nd round}
\label{fig:11}
\end{figure}

In the late 1970s, interest in topics of concurrency increased in the informatics community. In particular, process algebras such as calculus of communicating systems (CCS) and communicating sequential processes (CSP) emerged. However, semantics of such formalisms was undisputedly defined in terms of transition systems; viz., a single behavior (run) was -- and is -- conceived as a sequence of states and steps. Petri extended his concept of distributed runs as in Section~\ref{sec:3} \cite{Petri_79}. As an example, the above Figures~\ref{fig:03} and \ref{fig:05} show a run of Figure~\ref{fig:08}. Its final marking, $M_4$, enables the transition bake a second time.  Here the quest arises, where in the run of Figure~\ref{fig:03} this event is to be inserted (c.f. Figures~\ref{fig:09}): before move to shop? Together with move to shop? After move to shop but before sell? Together with sell? After sell? Are there five alternatives, depending on a shared clock? Do we assume the existence of a notion of “time”, in absolute precision, such that one of the five cases is “clearly taken”? Who decides this? What is assumed about “time” in Figure~\ref{fig:08}? Figure~\ref{fig:10} shows Petri’s answer: the event of 2nd baking occurs after the first bake and the supply to aide event, but independent of move to shop and sell. In technical terms, the second bake event is not ordered with respect to move to shop and sell. Notice that this kind of unorder can not be interpreted as “at the same time”. The events move to shop, sell and the second bake, cannot be totally ordered in any temporal sense. This is what the system in Figure~\ref{fig:08} expresses. Figure~\ref{fig:11} extends Figure~\ref{fig:10}: the second occurrences of move to shop and sell are inserted.

Yet, the idea of partially ordered runs never gained too much attention. They appeared intuitively odd, technically intricate, and overall not too beneficial. We return to these aspects later.

\section{The early 1980s}\label{sec:6}

Black token Petri nets are used to describe control and synchronization, and to count resources. In order to cope with data, it suggests itself to replace black dots by any kind of individual tokens, and to equip them with any kind of semantics. Various such versions of Petri nets have been suggested; essentially, two prevailed: colored Petri nets \cite{Jensen_Kristensen_09} and predicate transition nets \cite{Genrich_Lautenbach_81}. Colored net essentially equip Petri nets with conventional data structures as well-known from programming languages. Additionally, each data set is canonically expanded to the corresponding multiset.

\begin{figure}[tb]
\centering
\includegraphics[scale=.4]{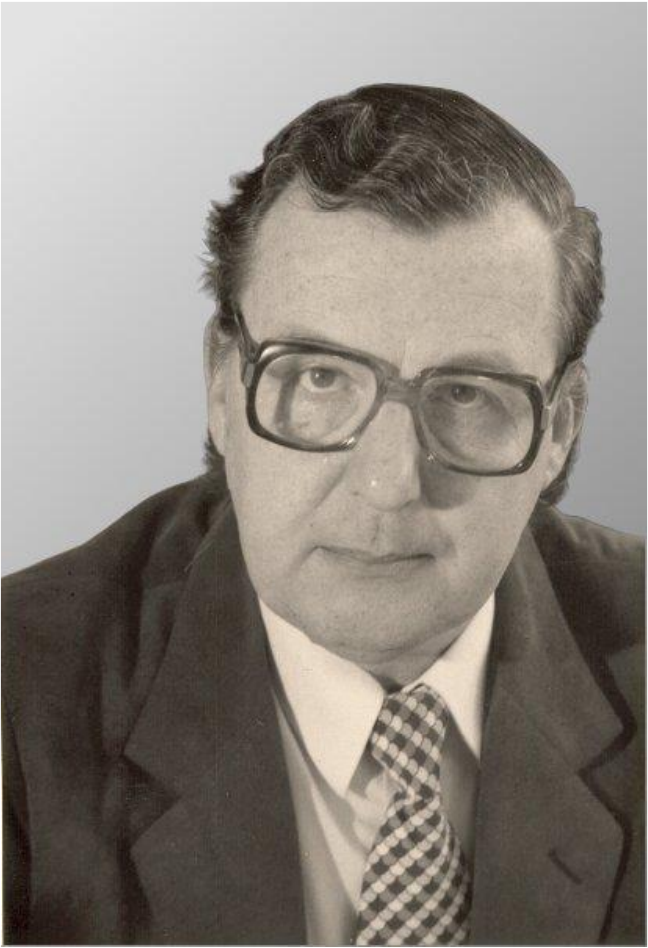}
\caption{Carl Adam Petri in the 1980s}
\label{fig:12}
\end{figure}

\begin{figure}[H]
\centering
\includegraphics[scale=.4]{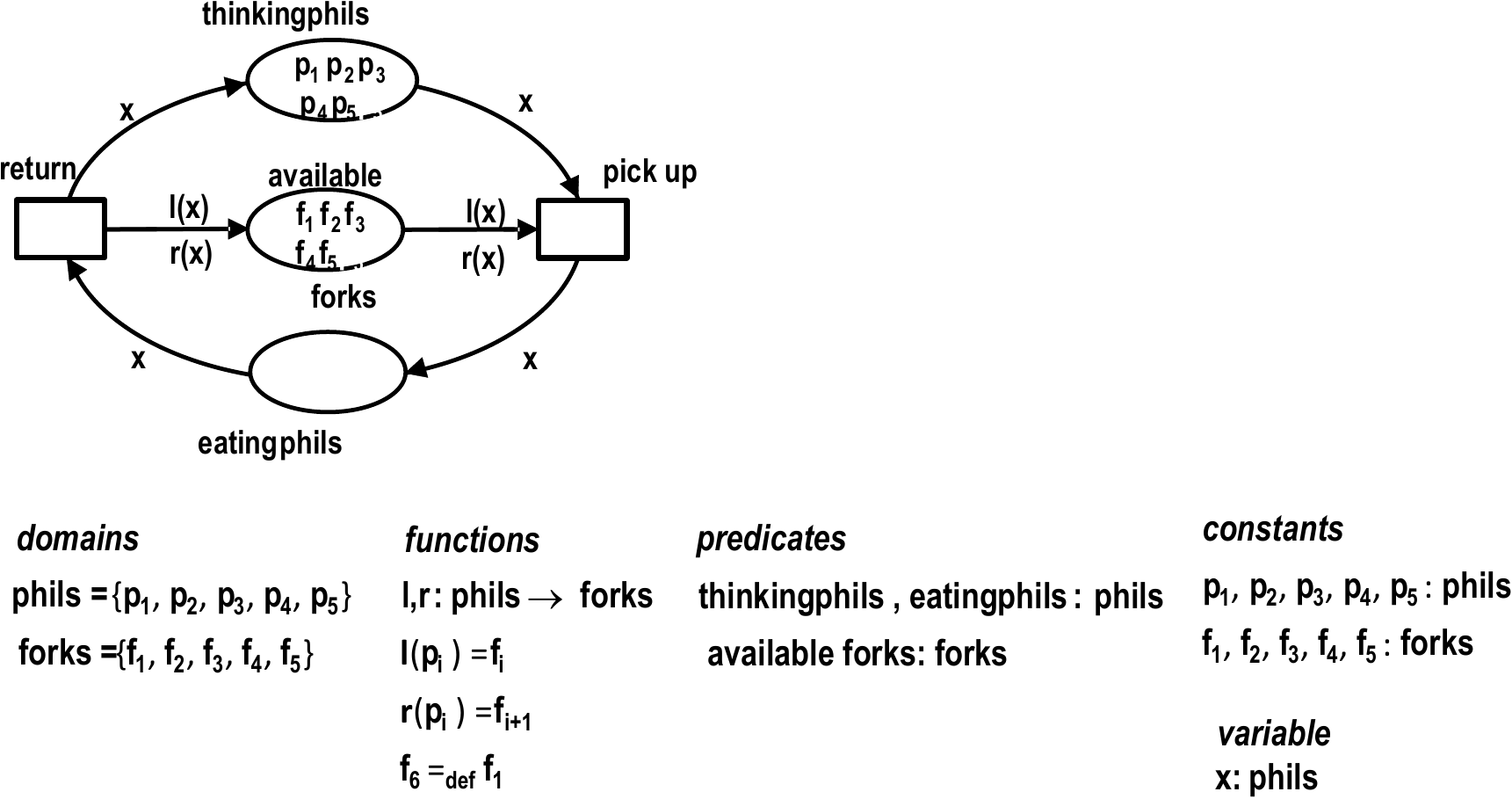}
\caption{Five thinking philosophers}
\label{fig:13}
\end{figure}

\begin{figure}[H]
\centering
\includegraphics[scale=.4]{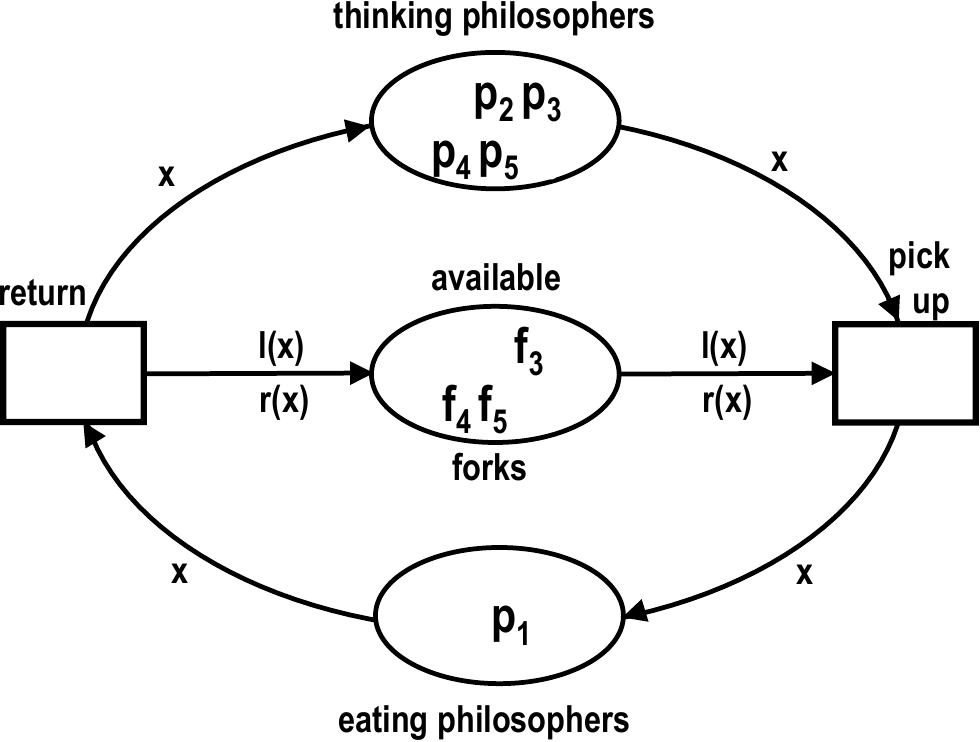}
\caption{Thinking philosophers: $x = p_1$}
\label{fig:14}
\end{figure}

\begin{figure}[H]
\centering
\includegraphics[scale=.4]{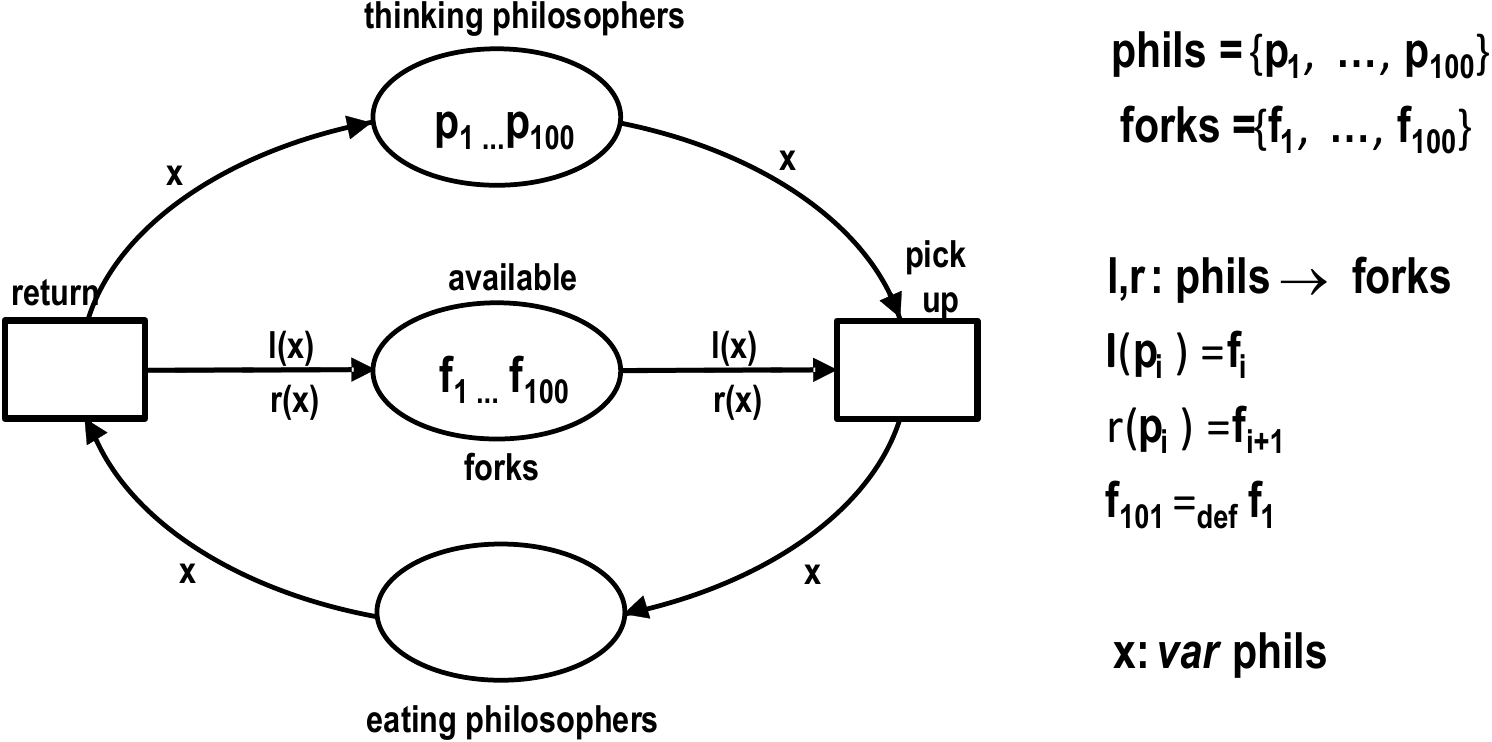}
\caption{$100$ thinking philosophers}
\label{fig:15}
\end{figure}

Section~\ref{sec:3} described the rise of Petri nets out of propositional logic. As the name suggests, predicate transition nets lifted this idea from propositional logic to predicate logic, extended by multisets. The basic idea is simple: Each place denotes a predicate $p$, and the place contains an element $a$, if $p$ applies to $a$. Figure~\ref{fig:13} shows an example: in the given situation, the predicate thinking phils applies to $p_1, \dots, p_4$, the predicate available forks applies to $f_1, \dots, f_5$, and the predicate eating phils applies to no element. Transition pick up is enabled for each valuation of the variable $x$ by one of the philosophers. With $x = p_1$, occurrence of pick up yields the situation of Figure~\ref{fig:14}. Finally, Figure~\ref{fig:15} shows the case of $100$ philosophers in a circle.

\begin{figure}[H]
\centering
\includegraphics[scale=.4]{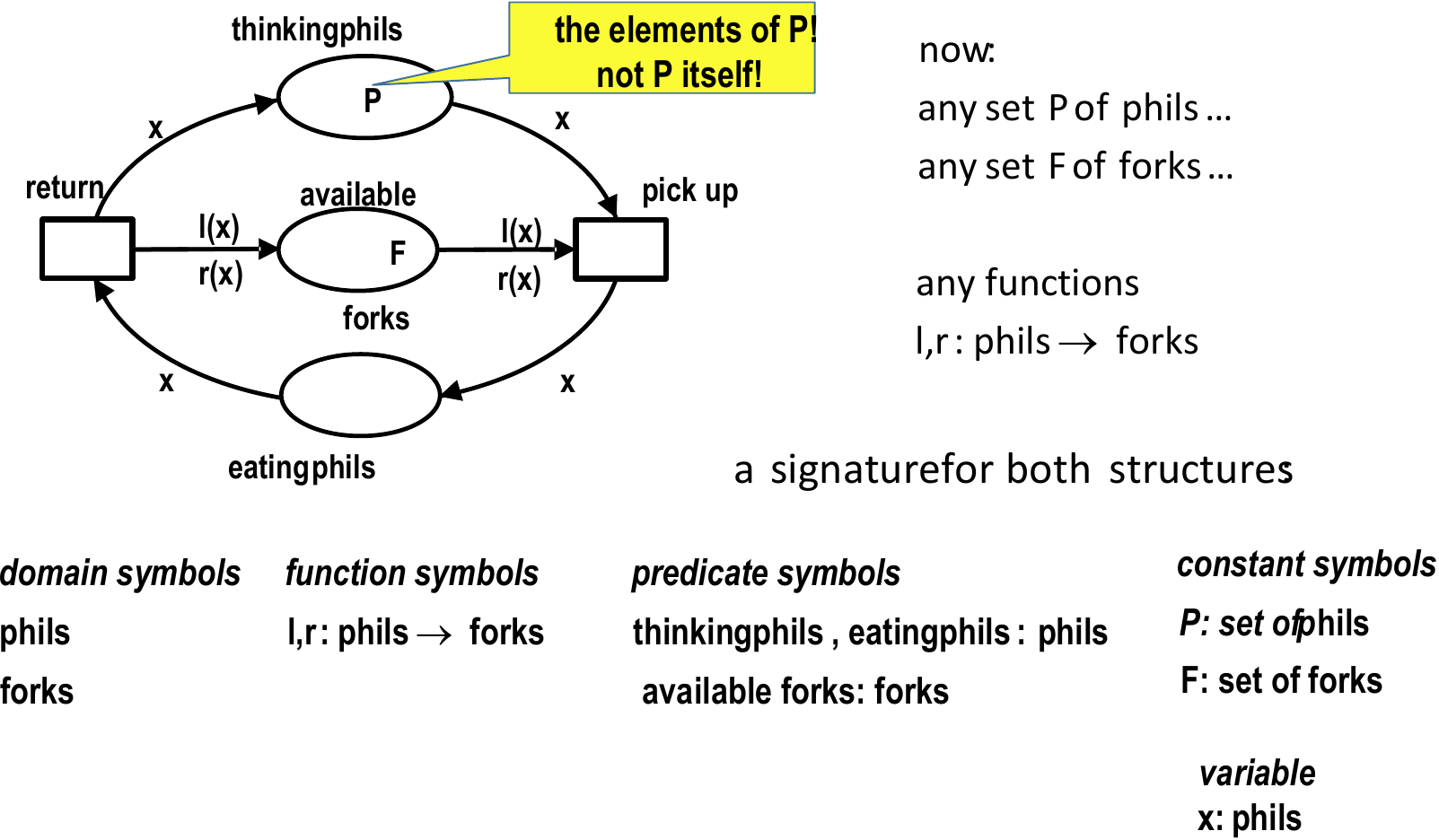}
\caption{Attempt}
\label{fig:16}
\end{figure}

Predicate logic is about symbolic, schematic representations. Real world items are then conceived as interpretations of symbols. As a first example, in Figure~\ref{fig:16}, the symbols $P$, $F$, $l$, and $r$ are intended to be interpreted by sets of philosophers and forks, and by corresponding functions. One such intended interpretation yields the system in Figure~\ref{fig:13}. Figure~\ref{fig:15} gives a second intended interpretation. However, Figure~\ref{fig:16} is fundamentally flawed: according to the occurrence rule of Petri nets, $P$ and $F$ are two tokens, albeit interpreted as sets of philosophers. Occurrence of pick up would remove the entire sets. But this is not intended: rather, single elements of these sets are to be removed: the symbol $P$ should not be interpreted as one set, but as “all single elements” of one set.

In the early 1980s, this problem has not been addressed. Only recently, in \cite{Fettke_Reisig_24}, it was suggested to understand this feature in the framework of predicate logic: Remembering that thinking phils and available forks are predicates and that the symbols $P$ and $F$ are to be interpreted as a set of philosophers and of forks, respectively, we want to express

\begin{equation}
\text{\say{$\forall p \in P: \sthinkingphilosophers(p)$}},
\end{equation}
\begin{equation*}
\text{and}
\end{equation*}
\begin{equation}
\text{\say{$\forall f \in F: \savailableforks(f)$}}.
\end{equation}

\begin{figure}[H]
\centering
\includegraphics[scale=.4]{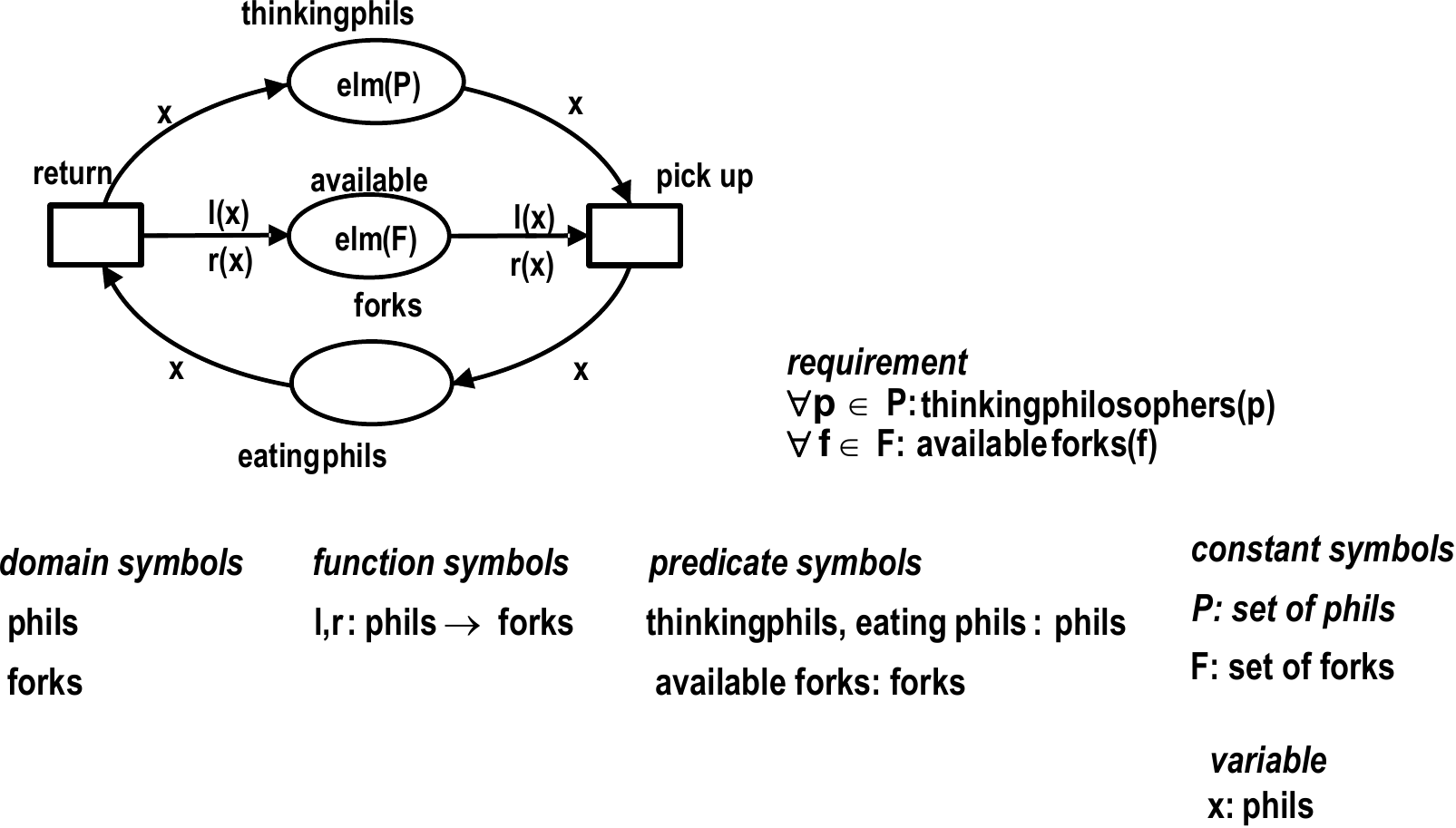}
\caption{Use of the $\elm$-operator}
\label{fig:17}
\end{figure}

We express this by the inscriptions

\begin{equation}
\elm(P) \text{ and} \elm(F),
\end{equation}
as in Figure~\ref{fig:17}.

\begin{figure}[H]
\centering
\includegraphics[scale=.4]{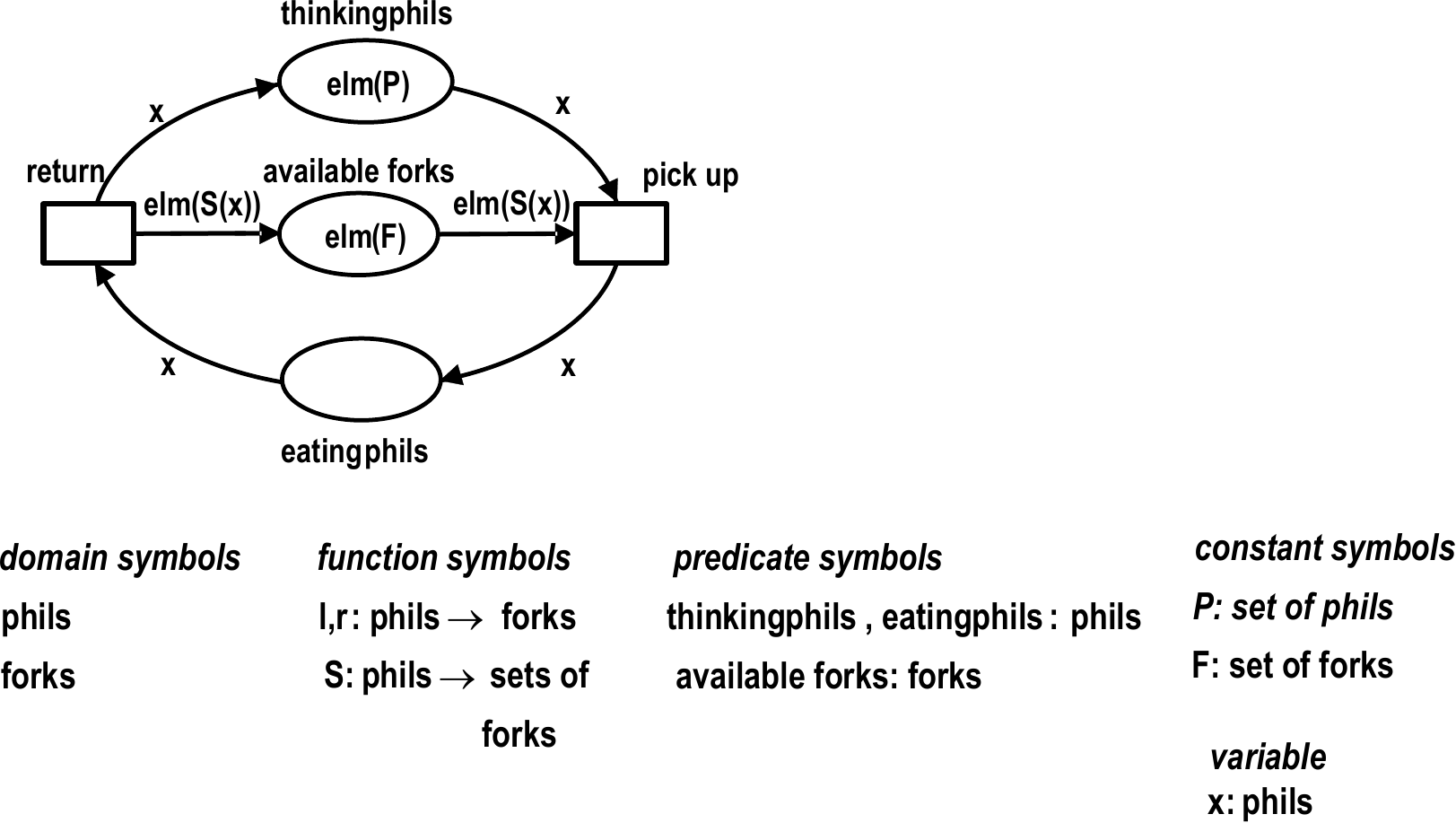}
\caption{Set of forks}
\label{fig:18}
\end{figure}

\begin{figure}[H]
\centering
\includegraphics[scale=.4]{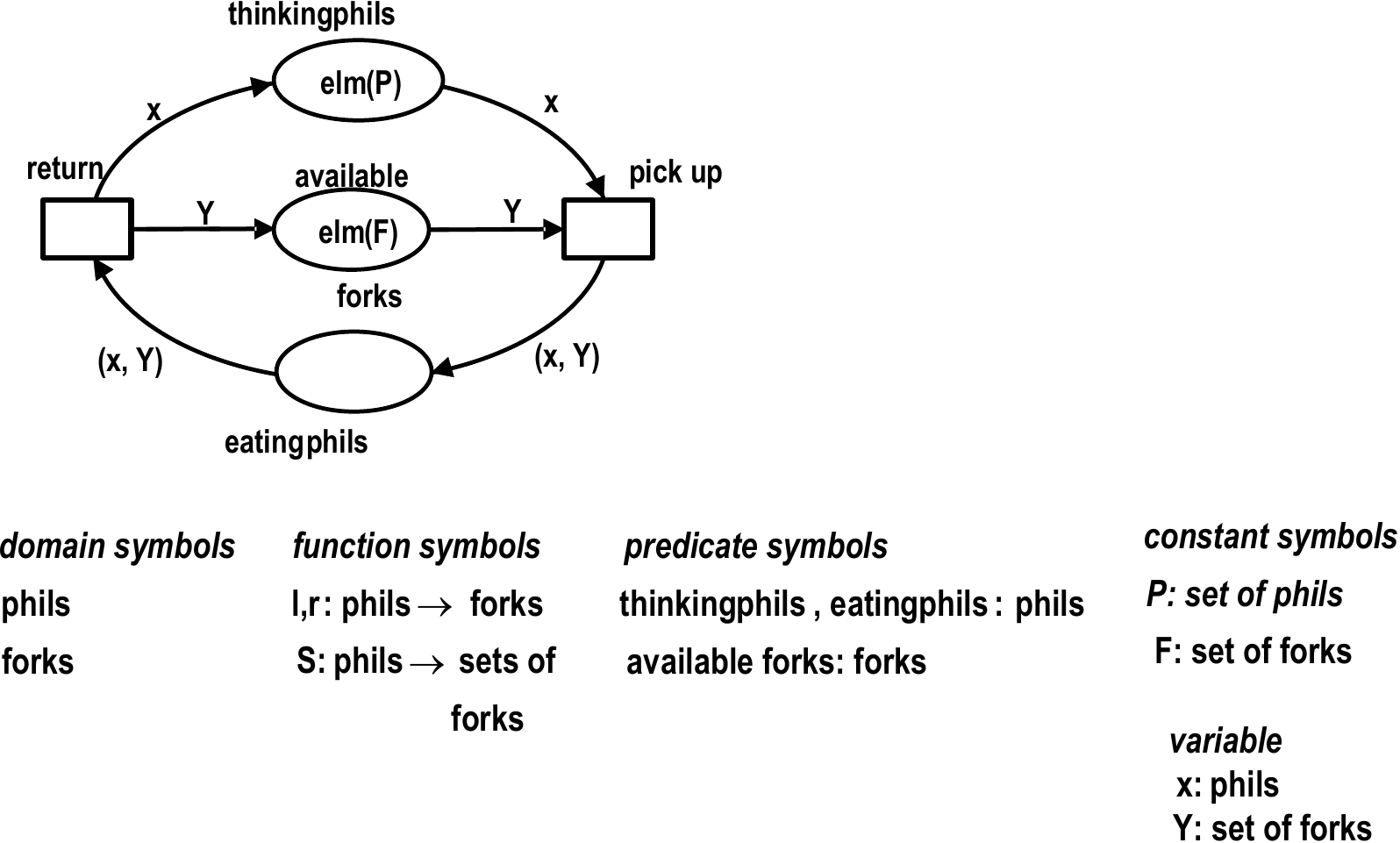}
\caption{Most liberal}
\label{fig:19}
\end{figure}

For example, each philosopher $p$ may be assigned its specific set $S(p)$ of forks, that $p$ picks up and returns upon occurrence of the transitions pick up and return in the mode $x = p$. Figure~\ref{fig:18} represents this case. Finally, each time forks are picked up, the set of picked up forks may change. Figure~\ref{fig:19} shows this case. More on the elm operator can be found in \cite{Fettke_Reisig_24}.

\begin{figure}[H]
\centering
\includegraphics[scale=.4]{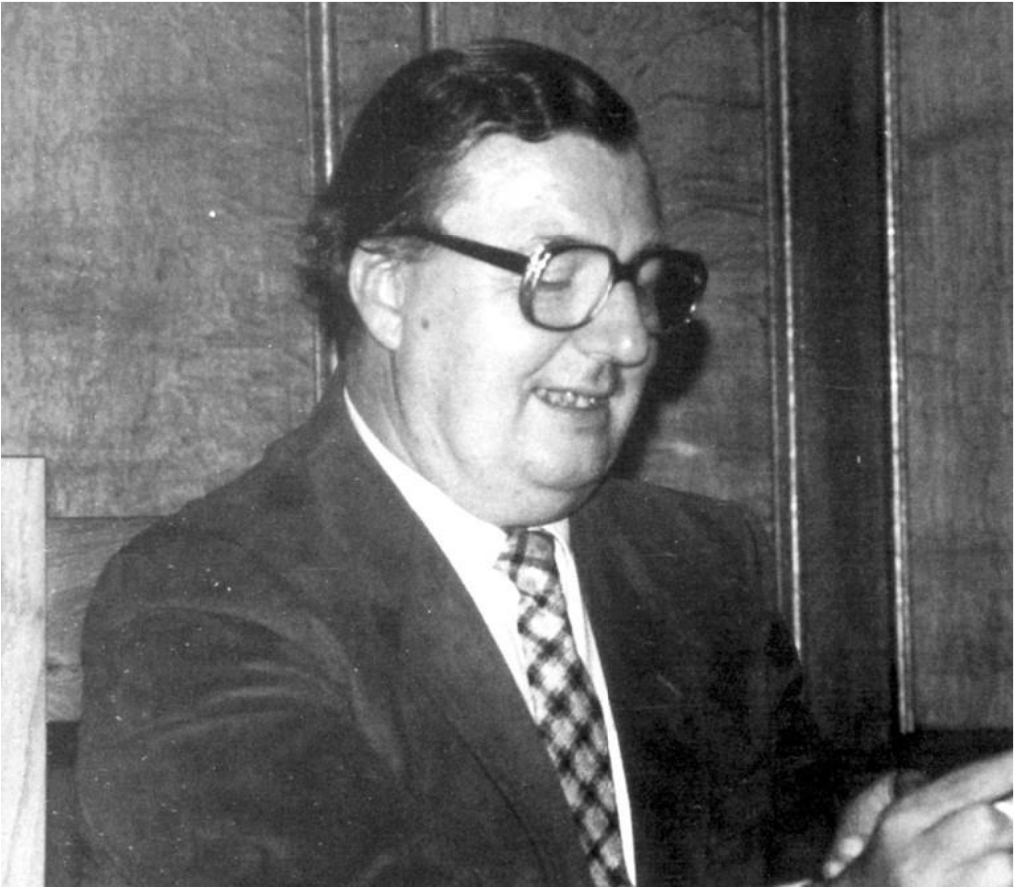}
\caption{Carl Adam Petri in the 1990s}
\label{fig:20}
\end{figure}

\section{Actual contribution: the composition problem}\label{sec:7}

Among long standing and actual problems and contributions to Petri nets, composition of nets certainly belongs to the most crucial ones. Composition of Petri nets has a long tradition, with many contributions. We just mention some relevant ideas of the last three decades: \cite{Christensen_Petrucci_93} composes many Petri net modules in one go, merging all equally labeled places as well as transitions. A long-standing initiative with many variants is algebraic calculi for Petri nets, such as the box calculus and Petri net algebras in various forms \cite{Best_et_al_01}. These calculi define classes of nets inductively along various composition operators, merging equally labeled transitions. \cite{Baldan_08} defines composition $A \compose B$ of “composable” nets $A$ and $B$ in a categorical framework, with $A$ and $B$ sharing a common net fragment, $C$. The effect of composition on distributed runs is studied. \cite{Kindler_Petrucci_09} suggests a general framework for “modular” Petri nets, i.e. nets with features to compose a net with its environment. A module may occur in many instances. Petri nets with two-faced interfaces are discussed in \cite{Rathke_et_al_14,Sobocinski_10}. They suggest Petri nets with boundaries (PNB), with two composition operators. 

The infrastructure \Heraklit suggests a most liberal approach with two-faced interfaces and an associative composition operator. In particular, \Heraklit allows mixed interfaces, i.e. two nets may be composed along places and transitions in one go. For technical details we refer to \cite{Fettke_Reisig_24}. In the sequel, we present some intuitively obvious examples, that show the expressive power of the \Heraklit composition operator. 

\subsection{A producer/consumer system}
\begin{figure}[H]
\centering
\includegraphics[width=\textwidth]{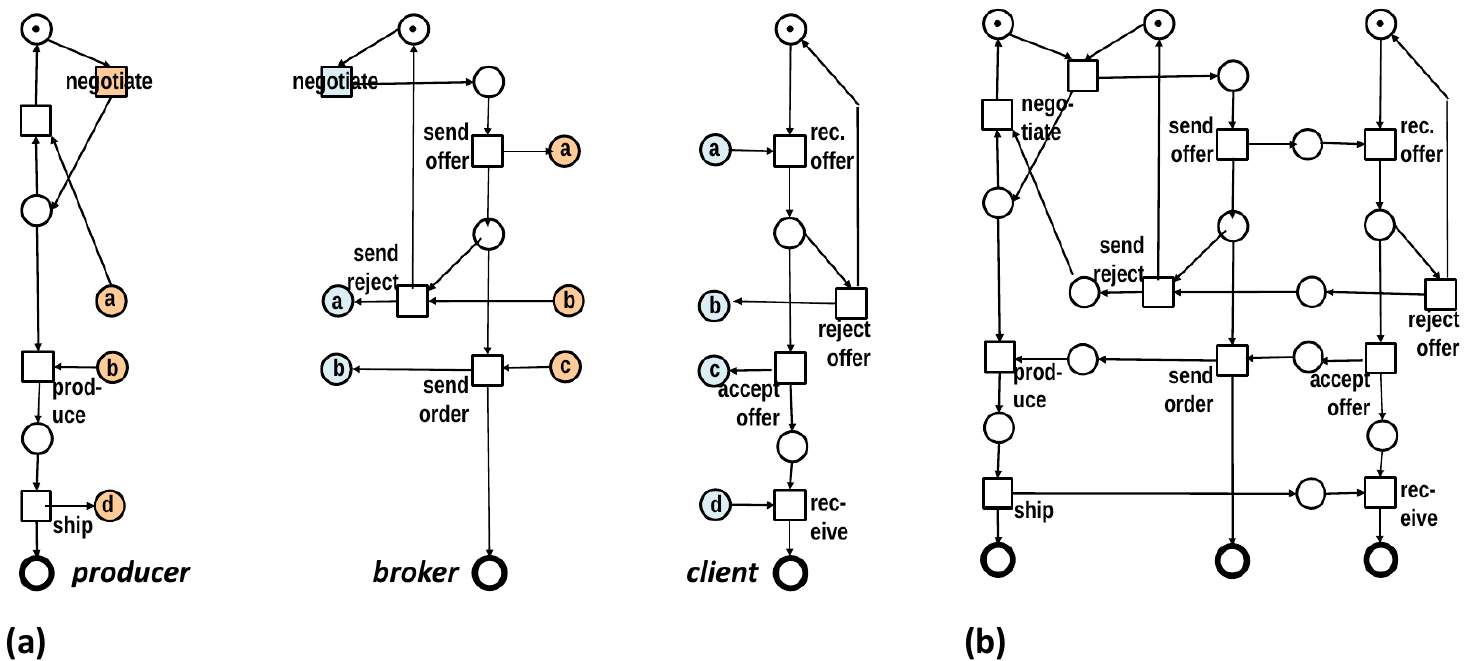}
\caption{(a) A producer, a consumer, and a client, and (b) the composition \textit{producer $\compose$ broker $\compose$ client}}
\label{fig:21}
\end{figure}

Figure~\ref{fig:21} shows three modules. The producer and the broker negotiate an offer to the client. If the client rejects, they re-negotiate the offer. If the client accepts, the producer produces a corresponding product and ships it directly to the client. Each module has a left and a right interface, with the corresponding elements drawn on the margin of the module’s surrounding box. The left interface of the producer, and the right interface of the client are empty. As a general rule, to compose two modules $A$ and $B$, equally labeled elements of the right interface $A^\ast $ of $A$ and the left interface $^\ast B$ of $B$ are merged. Figure~\ref{fig:21}(b) shows the composed module \emph{producer} $\compose$ \emph{broker} $\compose$ \emph{client}.

\subsection{Claim settlement}
\begin{figure}[H]
\centering
\includegraphics[scale=.6]{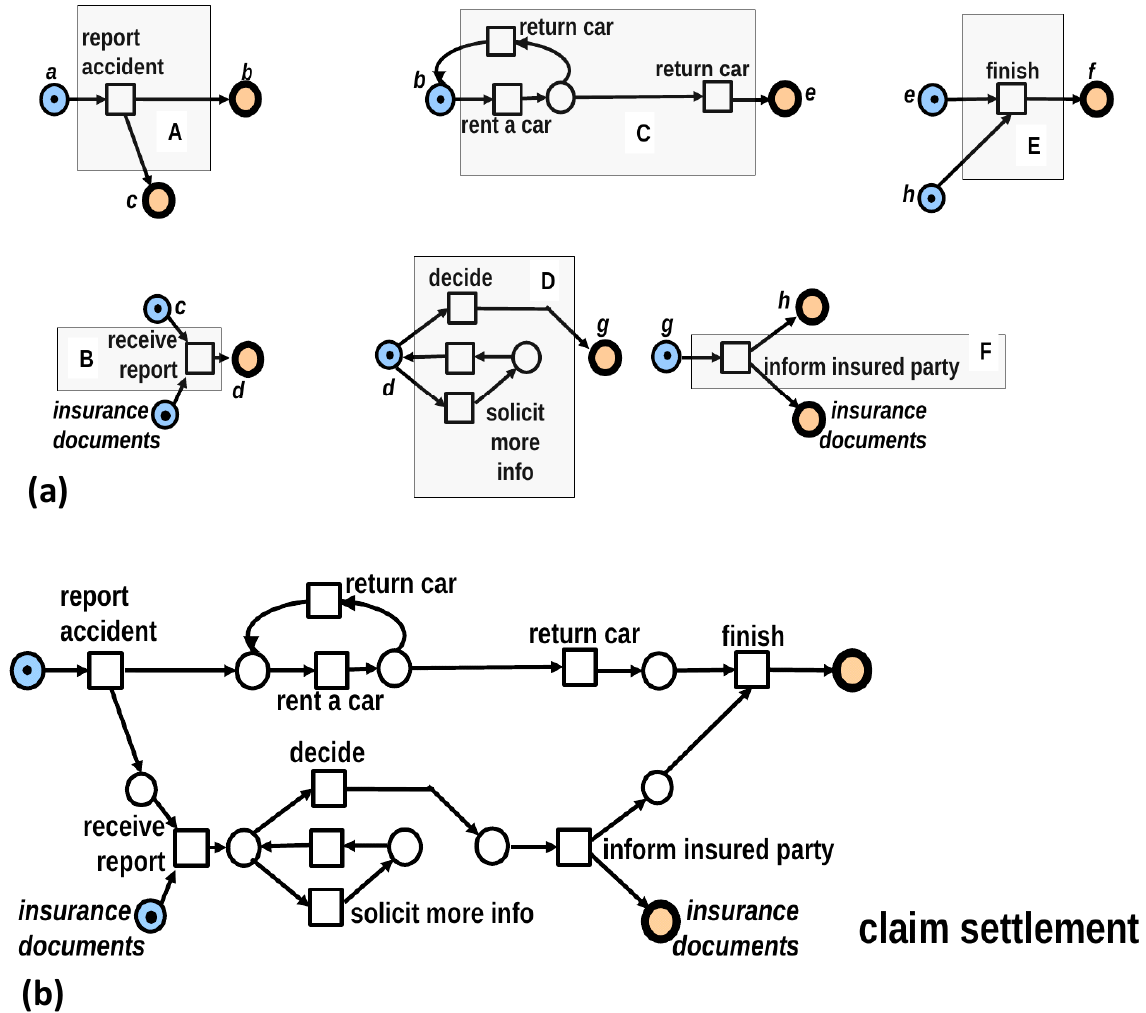}
\caption{(a) Six modules $A, B, C, D, E$, $F$, (b) their composition $A \compose B \compose C \compose D \compose E \compose F$}
\label{fig:22}
\end{figure}

As a second example, Figure~\ref{fig:22}(a) shows six modules, $A$, $B$, $C$, $D$, $E$, $F$, representing single activities of a car driver and his insurance company, in case of a car accident. Figure~\ref{fig:22}(b) shows the composition $A \compose B \compose C \compose D \compose E \compose F$, i.e. the overall behavior: the driver reports the accident and hires one or more times a car, until he is informed about the insurance’s decision. The insurance receives the driver’s report, may solicit more information, and eventually decides and informs the driver.

\begin{figure}[t]
\centering
\includegraphics[scale=.4]{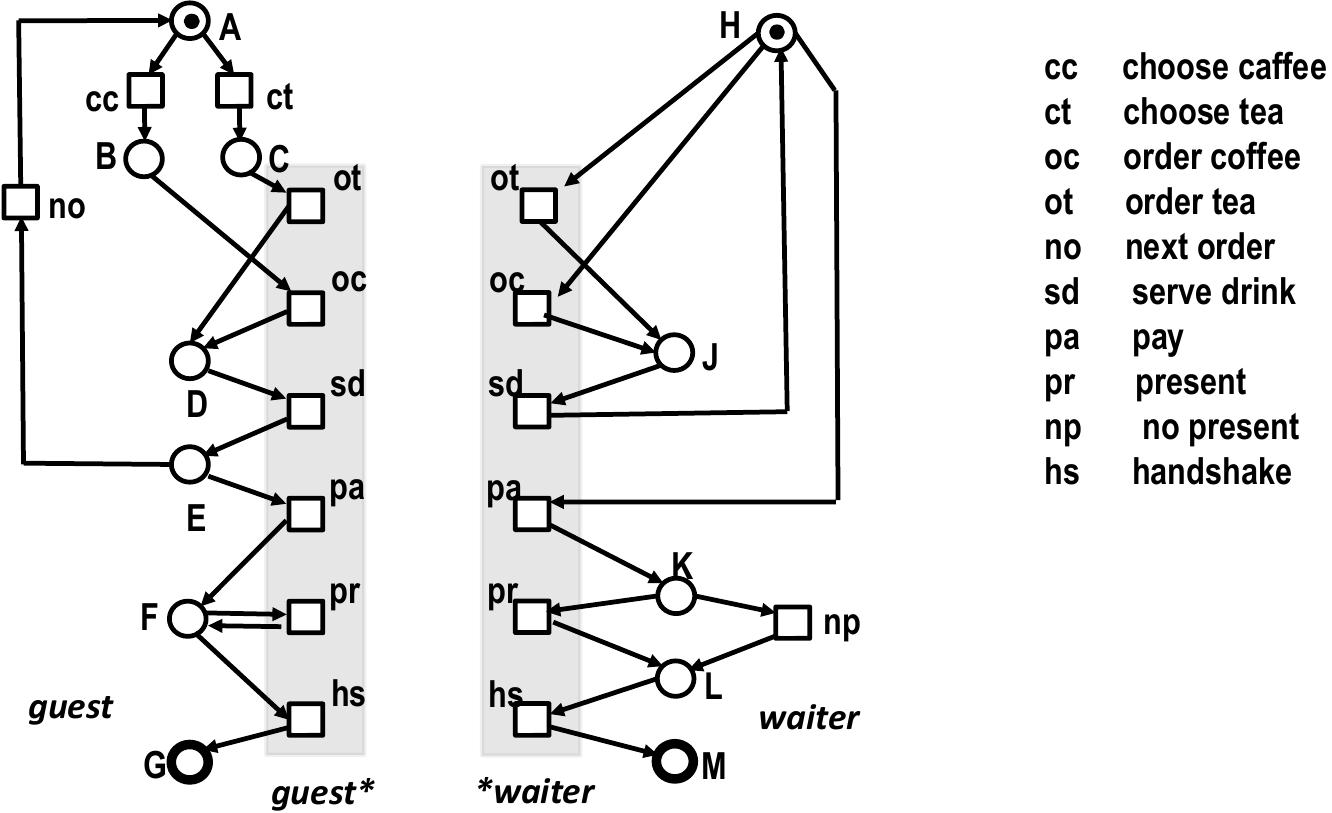}
\caption{Coffee house}
\label{fig:23}
\end{figure}

\subsection{Coffee house}
Figure~\ref{fig:23} shows the interaction of a guest and a waiter in a coffee house. Notice that the guest has a number of choices, to which the waiter reacts adequately. Composition $guest \compose waiter$ is obvious; we skip it here.

\subsection{The light/fan system}
A fundamental feature of Petri nets is the strict distinction between passive and active elements. It is often convenient to start modeling by identifying typical propositions as places, and involved activities as transitions. Then, each activity is related to the involved propositions, thus generating a single step. These steps can be composed into typical distributed runs of the intended elementary system module. Furthermore, to construct a corresponding elementary system module $N$, all occurrences of a proposition in all steps are merged, generating a place of $N$. The transitions and arcs of $N$ are inherited from the steps.

As an example of this procedure, we start with a colloquial description of a system containing a fan and a light, as they are standard in modern bathrooms: In the case of the fan is off, when you turn on the light, after some time, the fan will start running. In this situation, if you turn off the light, the fan continues running for some time. Hence, in the case of the fan is off, when you turn the light on and off quickly, the fan will not start running at all. And in the case of the fan on, when you turn the light off and on quickly, the fan will run continuously. To model this system, we first extract the involved propositions from the description: fan is off, fan is on, light is off, and light is on.  Furthermore, we identify four activities: turn light on, turn light off, fan starts running, and fan stops running.

\begin{figure}[t]
\centering
\includegraphics[scale=.4]{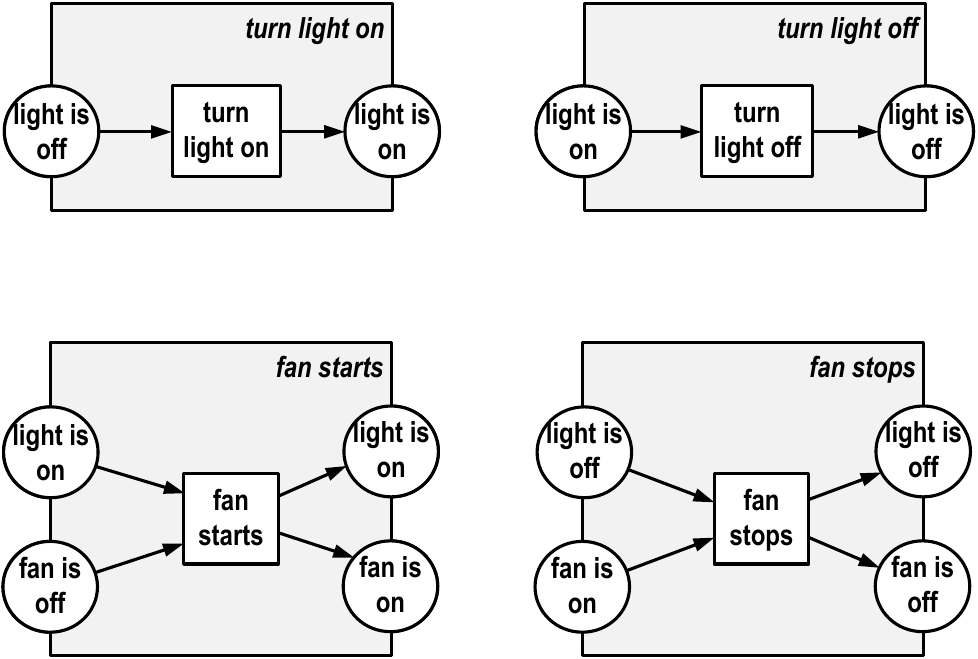}
\caption{Four steps}
\label{fig:24}
\end{figure}

\begin{figure}[t]
\centering
\includegraphics[scale=.4]{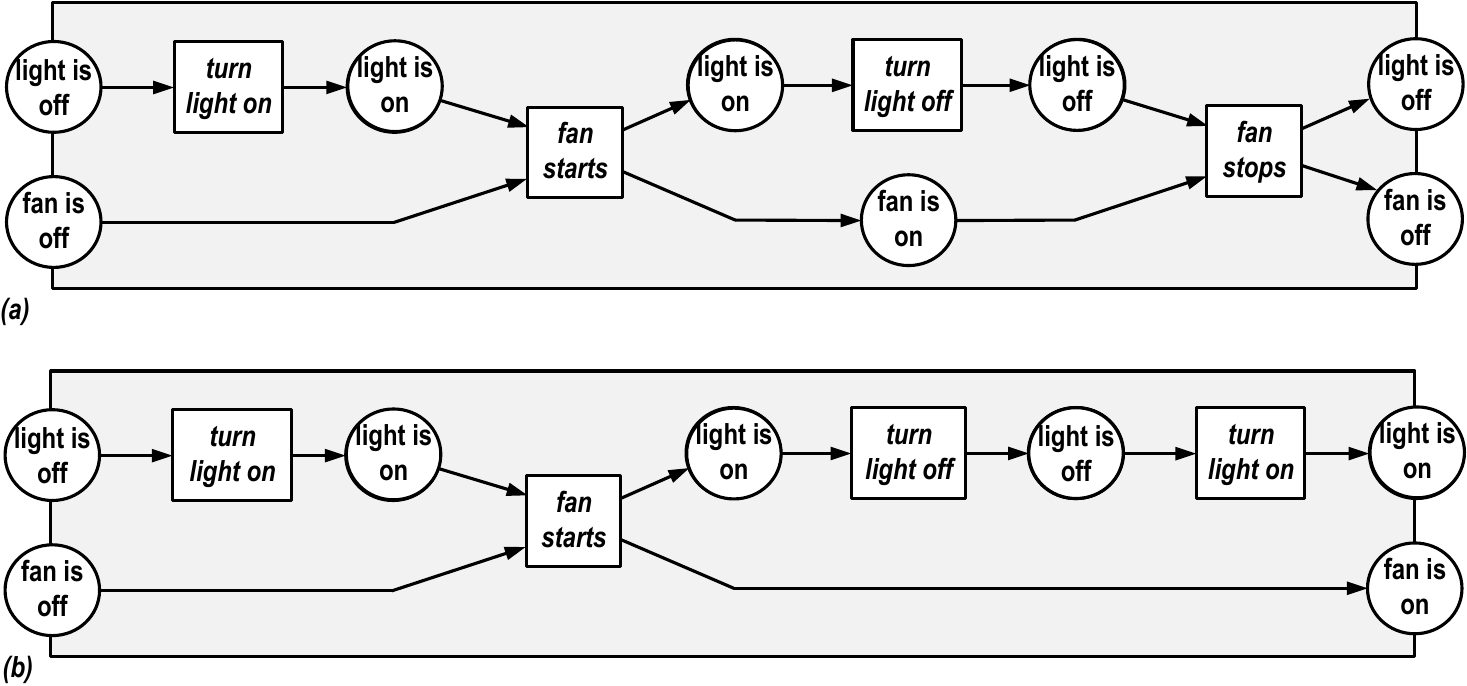}
\caption{Two runs: (a) \textit{
turn~light~on $\compose$ fan~starts $\compose$ turn~light~off $\compose$ fan~stops}, (b) \textit{
turn~light~on $\compose$ fan~starts $\compose$ turn~light~off $\compose$ turn~light~on}}
\label{fig:25}
\end{figure}

\begin{figure}[H]
\centering
\includegraphics[scale=.4]{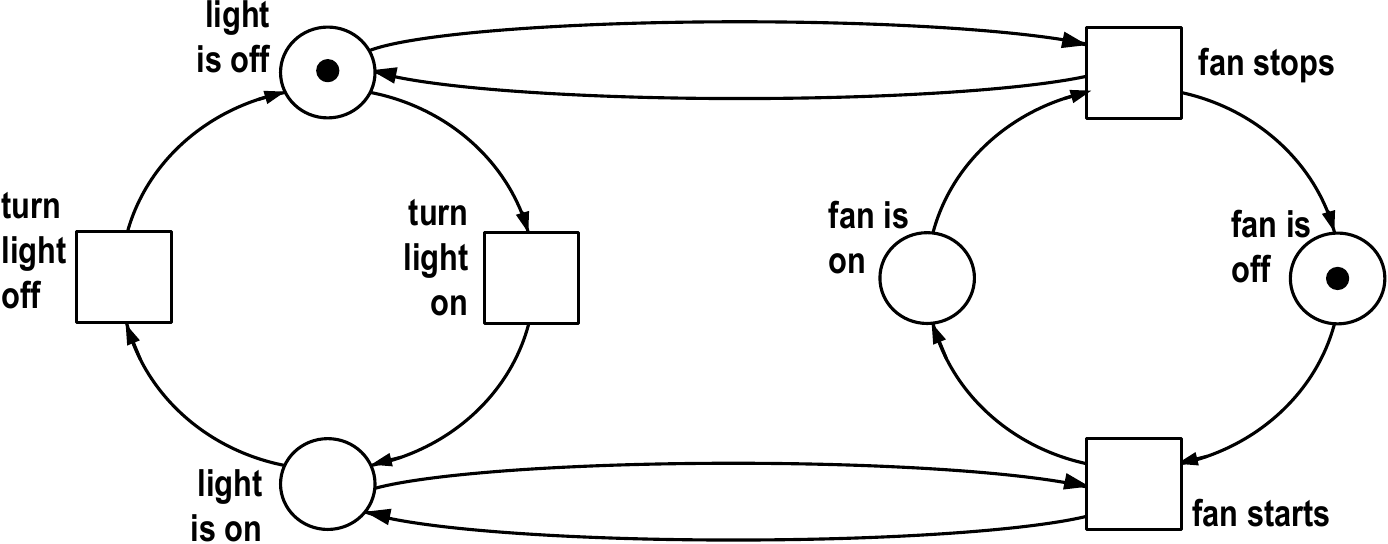}
\caption{The light/fan system}
\label{fig:26}
\end{figure}

The effect of each activity on the propositions can be represented as a step, shown in Figure~\ref{fig:24}. Starting in a situation with both light and fan off, the system can evolve in different runs; it is non-deterministic. Figure~\ref{fig:25} shows two such runs. The system itself permits infinitely many runs. It is now easy to derive an elementary system module $N$, such that the runs of $N$ are exactly the runs of the system: for each proposition $p$ in the steps of Figure~\ref{fig:24}, merge all occurrences of $p$ into one place of $N$. This results in the elementary system module of Figure~\ref{fig:26}. As an initially assumed state one may choose both the light and the fan off.

\section*{Conclusion}
The essence of Petri net theory, as Petri saw it, was a comprehensive look at informatics, that differs from that what in fact emerged in the 1960s. In his view, a theory of informatics

\begin{itemize}

\item is about discrete models of systems (including computers);

\item should base on a dynamization of logic (propositions and predicates with varying extensions);

\item respects local causes and effects of events, hence considers distributed runs;

\item avoids unbounded means to describe, to measure, to model systems. 
\end{itemize}
 
Consequently, a theory of informatics must not be about abstract, symbol crunching automata -- that per se cannot be implemented.

More on Carl Adam Petri as a person and more about his ideas can be found in \cite{Scmith_15}.


%
%
%

\bibliographystyle{splncs04}
\bibliography{main}

\end{document}